\documentclass[preprint,showpacs,showkeys,pra,letterpaper]{revtex4}
\usepackage{epsf}
\usepackage{epsfig}
\usepackage{amssymb}

\newcommand \be{\begin{equation}}
\newcommand \ba{\begin{eqnarray}}
\newcommand \ea{\end{eqnarray}}
\newcommand \ee{\end{equation}}

\begin{document}
\title{Waveguiding power of photonic crystal slabs}
\author{ Serge Luryi and Arsen V. Subashiev$^{*}$}
\affiliation{Department of Electrical and Computer Engineering,
\\ State University of New York at Stony Brook, Stony Brook, NY
11794-2350 \\
\small {\textit{$^{*}$Corresponding author:
Subashiev@ece.sunysb.edu}}}
\begin{abstract}
We consider the waveguiding by  thin patterned slabs embedded in a
homogeneous medium. In the longwave limit, the wave spectra of slabs
are found to be well described by a single frequency-independent
parameter, which we call the ``guiding power". The guiding power can
be evaluated in an effective medium approximation, similar to the
Maxwell Garnett theory, but modified for the local field corrections
specific to the two-dimensional geometry. The guiding power is
different for the transverse magnetic (TM) and transverse electric
(TE) polarizations. We show that the confinement factor of TM waves
in a porous layer with high index ratio can exceed that for a
homogeneous layer. Similarly enhanced confinement of TM waves is
demonstrated for a layer of elongated cylinders or elliptic
inclusion with a high axis length ratio. The effect originates from
the suppression of local field effects and the increasing internal
field in the inclusion. It may be useful in the design of
far-infrared or THz quantum cascade lasers.
\end{abstract}
 \maketitle

\section{Introduction. Guiding power}

Waveguiding of light in layered patterned structures, such as slabs
of two-dimensionally periodic photonic crystals (2D PC), has
attracted much interest in view of potential photonic applications
\cite{Yablonov,Painter,Joannop}. The studied 2D patterns include
periodic lattices of deep etched air pores or ``conjugate" lattices
of high permittivity cylinders. The patterned slabs (not necessarily
periodic) can be employed in waveguides as either a core or a
cladding. Numerous theoretical computations of the band spectra of
2D PC and PC slabs have been reported, based on expansions of the
electromagnetic field in plane waves \cite{Maradu,Joannop} or
cylindrical waves \cite{Nicoro,Ohtaka}, as well as based on
finite-difference time-domain methods \cite{TafLove,Borodit}.

The low-frequency region of electromagnetic waves in the 2D PC is
well understood. The waves have a linear spectrum that is very close
to that obtained in the effective media approximation
\cite{Sarychev,Halevi} with an effective permittivity corresponding
to the Maxwell Garnett theory \cite{Sarychev,Nicoro2}. This means
that when the wavelength $\lambda$ exceeds the structure period $a$,
the optical properties are primarily determined by the filling
factor $f$ of the inclusions (i.e. their total volume fraction) and
do not rely on their long-range order or their shape variation. The
disorder leads to a weak (for $\lambda \gg a$) Rayleigh-like
scattering.

In this paper we investigate the waveguiding by PC slabs in a long
wavelength frequency range, $qd \ll 1$, where $q$ is the wave
vector, $d$ is the thickness of the active or a core layer of the
waveguide. We show that in this range the waveguiding has a
universal form described by a single parameter, which we call the
``guiding power". For short-period structures, $d \gg a$, and for
sparse structures, $d \ll a$,  it can be calculated in terms of the
polarizability of the patterned dielectric core via a
self-consistent procedure to include local field effects.

We discuss the waveguiding in highly inhomogeneous structures,
such as planar regular arrangements of nearly overlapping cylindrical pores or
high-index cylindrical rods with large spacings. For thin slabs, the
local field effects are different from those in an infinitely
extended 2D PC, primarily because of the short-range dipolar
interaction between the finite-height cylinders or spheres. Proper
inclusion of the local field effects in the low-frequency region
enables a perturbative approach with fast convergence.

In the design of far-infrared and terahertz semiconductor lasers the
optical confinement is an important issue. It poses a severe problem
for quantum cascade lasers where intersubband radiative transitions
require transverse magnetic (TM) polarization of emitted waves
\cite{Capasso}. With this polarization in the long wavelength
region, the so-called modal confinement factor $\Gamma_{\rm TM}$  is
known to be small for any contrast of the dielectric constants
between the core and the claddings \cite{Marcuse,Visser}. The reason
for the small values of $\Gamma_{\rm TM}$ is the reduced electric
field in the high-index core due to the boundary conditions at the
core layer surface. The decrease of the electric field devalues
traditional attempts to improve waveguiding by choosing cladding
layers with lower index.

In this paper, we propose to enhance the confinement by using
penetration of the electric field into a patterned core. The model
core under consideration comprises additional cylinder inclusions of
high dielectric constant compared to that of the claddings. The
filling factor for these inclusions is small, so as not to disturb
processes in the active region (which is the remainder of the core
outside the inclusions). We show that the better penetration of the
field can result in a much stronger waveguiding (much reduced field
spread outside the layer), as compared to the homogeneous layer.
This case has an advantage of not being critically sensitive to the
composition of the claddings.

An alternative way to improve the guiding of TM waves is to use a
porous PC slab  as the active layer. We show that the confinement of
TM modes in a structure with a high index contrast (typical for
silicon-on-insulator devices) can be enlarged if the active layer
has a patterned structure with pores. Even though the average index
of the porous core is reduced, this is more than compensated by
better penetration of the electric field into the structure, so that
the wave confinement is ultimately enhanced.

Patterned structures with the enhanced confinement can be
advantageously used in the laser design to minimize the losses from
free-carrier absorption and reduce the threshold current.

An unusually strong guiding of TM waves in a layer of cylinders was
previously observed in numerical studies of waveguiding by PC slab
structures \cite{Johnson1,Johnson2} but the effect had not been
properly recognized or explained.

\section{Weakly guided waves in laterally uniform waveguides}
In order to introduce the concept of weak waveguiding, let us first
consider the propagation of an electromagnetic wave along a
laterally uniform dielectric waveguide with a core layer. Let the
index profile depend only on $z$, approaching at large $|z|$ the
(background) dielectric constant $\epsilon_b$ of the cladding
layers.

Two wave polarizations are distinguished by the field orientation
relative to the structure symmetry plane. Consider the case of a TE
mode, when the electric field has only an in-plane $y$ component and
is strictly perpendicular to the wave propagation direction $x$. The
wave equation for the electric field $E_y=E_y(z)\exp(iqx)$ is of the
form
\begin{equation}
{{d^2} \over {dz^2}} E_y(z) =  [q^2  - k_0^2 \epsilon(z)] E_y ~,
\label{App-1}
\end{equation}
where $\bf q$ is the 2D wave vector in the plane of the waveguide.
Let us integrate Eq. (\ref{App-1}) between $-z_1$ and $z_1$, that is
over the region where the permittivity is variable. At $|z| \ge z_1$
the solution of the Eq. (\ref{App-1}) has the form
$E_y(z)=E(0)\exp(\pm \kappa z)$ and we get
\begin{equation}
- 2 \kappa  E_y(0) = \int_{-z_1}^{z_1} [q^2  - k_0^2 \epsilon(z)]
E_y(z)dz ~. \label{App-2}
\end{equation}
In the limit of weak waveguiding, $\kappa z_1 \ll 1$, the field
$E_y(z)$ is a slowly varying function across the entire layer.
Therefore it can be replaced in the integral by a constant value
taken, e.g., at $z=0$. Outside the guiding layer we have $q^2=
\epsilon_b k_0^2+\kappa^2$ and hence in Eq. (\ref{App-2}) we can
take $q^2=\epsilon_b k_0^2$ as the zeroth-order approximation. This
yields
\begin{equation}
 \kappa  ={1 \over 2} \epsilon_b k_0^2 \int_{-\infty}^{\infty} \left(
{\epsilon(z)\over\epsilon_b}-1\right)dz ~. \label{App-3}
\end{equation}
We have replaced the limits of integration by $\pm \infty$, since
the region where $\epsilon(z)=\epsilon_b$ does not contribute to the
integral in Eq. (\ref{App-3}).

Similar arguments can be used to consider the waveguiding of TM
waves, when the only non-vanishing component of the magnetic field
is $H_y$. For this case, in the long wavelength limit, the quantity
that remains a smoothly varying function across the layer is the
normal component of the electric displacement vector $D_z$.
Integrating the wave equation for $D_z$ we obtain
\begin{equation}
 \kappa   = {1 \over 2 }\epsilon_b k_0^2 \int_{-\infty}^{\infty}
 \left(1 -{\epsilon_b \over \epsilon(z)}\right) dz
 . \label{App-6}
\end{equation}
For both polarizations the dispersion relation for the guided wave
is of the form
\begin{equation}
\epsilon_bk_0^2=q^2 -\kappa^2 , \label{disp-main}
\end{equation}
where $k_0=\omega/c$ is the frequency parameter. Parameter $\kappa$
describing the exponential decay of the wave away from the core,
$\exp (-\kappa z)$, depends on frequency. Since $\kappa \propto
k_0^2$ the spectrum of the guided wave in the long-wavelength limit
has a universal character. It is convenient to introduce another
parameter $g$,
\begin{equation}
\kappa ={1 \over 2} \epsilon_b k_0^2 g , \label{kappa}
\end{equation}
which we shall call the ``guiding power" of the high-index core. The
value of $g$ defined by Eq. (\ref{kappa}) is owing to the fact that
it is frequency-independent in the weak guiding limit, $\kappa d \ll
1$ \cite{Marcuse}. So long as $g$ is constant, Eqs.
(\ref{disp-main}, \ref{kappa}) define a universal dispersion
relation for the guided modes of a three-layer dielectric waveguide
of core thickness $d$.

According to Eqs. (\ref {App-3}) and (\ref {App-6}), the guiding
power is given by :
\begin{equation}
g_{\rm {TE}} = \int_{-\infty}^{\infty}
\left ( {\epsilon(z)\over\epsilon_b}-1\right)dz~,~~~
g_{\rm {TM}} =  \int_{-\infty}^{\infty}  \left(1 -
{\epsilon_b \over \epsilon(z)}\right) dz
\label{multilayer_wg}~.
\end{equation}
For the simplest case of a constant permittivity ($\epsilon_g$)
core layer, Eqs. (\ref {multilayer_wg}) reduce to
\begin{equation}
g_{\rm TE} ={{\epsilon_g-\epsilon_b} \over {\epsilon_b}} d~,~~~
g_{\rm TM} ={{\epsilon_g-\epsilon_b} \over {\epsilon_g}} d~.
\label{TM-film}
\end{equation}

In structures with a low index contrast, $(\epsilon_g  -
\epsilon_b)\ll  \epsilon_b$, the values of $g$ for both modes are
small and close to each other. In the opposite limit, $(\epsilon_g /
\epsilon_b)\gg 1$, the guiding power for the TM mode is ${\epsilon_b
/ \epsilon_g}$ times weaker than $g_{\rm TE}$, which can be
explained by the reduced $z$ component of the electric field inside
the slab.

Confinement of guided waves is usually described by the
dimensionless ``confinement factor" $\Gamma$ (fraction of the wave
intensity that flows in the high-index core). Quite generally,
$\Gamma$  is proportional to the guiding power. The condition
$\Gamma \ll 1$ corresponds to the {\em weak guiding limit}. In this
limit the guiding power determines both the confinement properties
and the dispersion of waves.

\section{Waveguiding by a PC slab}
Consider now the electromagnetic wave propagation along a photonic
crystal slab formed either by a lattice of holes in a core layer or
by a ``mirror" structure with a set of cylinder rods serving as the
core. In the long wavelength limit $\lambda \gg a, d$, where $a$ is
the PC lattice constant and $d$ is the layer thickness, the field
inhomogeneity is important only at short distances away from the
slab, since the short-range components of the fields decay
exponentially over the distances of order $a$. This allows us to
identify the polarized waves as TE-like and TM-like. For $\lambda >
d$ the weakly bound guided waves have a smooth exponential decay of
the wave field away from the core, $E_z=E_{z,~\rm{out}}\exp(-\kappa
z)$ with $\kappa d \ll 1$. The relationship between $\kappa$ and $q$
can be obtained either by integrating the wave equation using the
weak-guiding approximation (as in the preceding section) with
averaging in the lateral plane, or by using the effective media
approach. The latter corresponds to replacing the PC layer by a
homogeneous slab with an effective (anisotropic) dielectric constant
\cite{ASSL}. Equations (\ref{disp-main}, \ref{kappa}) remain valid
in both cases, but the guiding power $g$ must now be evaluated
taking account of the polarizability of the laterally inhomogeneous
guiding layer.

Consider a core comprising a set of dielectric cylinders of radius
$r$ with the dielectric constant $\epsilon_{cyl}$  and height $d\gg
r$. The cylinders are spaced apart with a lattice constant $a$ and
the core medium between the cylinders is assumed to have the same
permittivity $\epsilon_b$ as the background. To compare the results
with those of full-scale calculations, we further consider a square
lattice of cylinders and the structure parameters close to those
studied in \cite{Johnson1,Johnson2}.

Importantly, the polarizability of a single cylinder is highly
anisotropic. For a sufficiently elongated cylinder, $d/r \ge 5$, the
polarization vector inside the cylinder is homogeneous and equal to
that of an ellipsoid with a high axes length ratio. Besides, one
must allow for depolarization effects.  We take them into account
approximately, by replacing the cylinders by prolate ellipsoids with
the same diameter and the same volume, so that the axes length ratio
of the ellipsoid $R_e=3/4 (d/r)$. The dipole moment of a single
cylinder in an external electric field equals $P_z = V \alpha_z
E_z$, where $V=\pi r^2d~$ is the cylinder volume and the
polarizability of a single cylinder  $\alpha_z$ is given by
\be \alpha_z =\frac{1}{4\pi}    \epsilon_b\frac{\epsilon_{\rm cyl}
-\epsilon_b} {\epsilon_b +(\epsilon_{\rm cyl} -\epsilon_b)n_z}~.
\label{PolCyl} \ee
Here $n_z$ is the depolarization factor \cite{L&L8}. Even though in
our case $n_z\ll 1$, the product $\epsilon_{\rm cyl~}n_z$ cannot be
neglected. The average dielectric function of the layer equals
\be \epsilon_{l,~\|}=\epsilon_b+ f \epsilon_b \frac{{\epsilon_{\rm
cyl} -\epsilon_b} }{\epsilon_b +(\epsilon_{\rm cyl}
-\epsilon_b)n_z}~. \label{EpsAv} \ee
The guiding power can now be calculated as in Eq. (\ref{TM-film}),
\be g_{\rm TM}=d(1-\epsilon_b/\epsilon_{l,~\|})~, \label{gptm0} \ee
and is given by \be g_{\rm TM, ~\rm cyl}=d\frac{f(\epsilon_{\rm cyl}
-\epsilon_{b})}{\epsilon_{b}+(f+n_z)(\epsilon_{\rm cyl}
-\epsilon_{b})}~. \label{gptm1}\ee
The spectrum of the TE mode can be calculated in a similar way, with
the transverse layer polarizability, $\epsilon_{l,~\bot}$, expressed
in terms of the transverse polarizability of a single cylinder,
$\alpha_y$. The result is similar to Eq. (\ref{PolCyl}) except for
the depolarization factor, which should now be replaced by $n_y$.
For a highly elongated cylinder, $n_y$ is very close to 0.5, so that
\be
\alpha_y=\frac{1}{2\pi}  \epsilon_{b}\frac{\epsilon_{\rm cyl} -\epsilon_{b}}  {\epsilon_{\rm
cyl}+\epsilon_{b} }~. \label{EpsBot} \ee
In calculating the average dielectric function $\epsilon_{l,~\bot}$
of the layer one should take into account the local field effects
(enhancement of the local field due to the field of surrounding
cylinders). The simplest way to do this is to use the Maxwell
Garnett approach, which is strictly applicable for $d\ll a$. In this
approximation we have
\be \epsilon_{l,~\bot}= \epsilon_{b}\frac{
\epsilon_{b}+\epsilon_{\rm cyl}+ f (\epsilon_{\rm cyl}
-\epsilon_{b})}{\epsilon_{b} +\epsilon_{\rm cyl}-f(\epsilon_{\rm
cyl} -\epsilon_{b})}~. \label{EpsBot0} \ee
The guiding power for TE waves is then given by
\be g_{\rm TE, ~\rm cyl}=2d\frac{f(\epsilon_{\rm cyl}
-\epsilon_{b})}{\epsilon_{b}+\epsilon_{\rm cyl}-f(\epsilon_{\rm cyl}
-\epsilon_{b})}~. \label{gpte1} \ee
\begin{figure*}[t]
\epsfig{figure=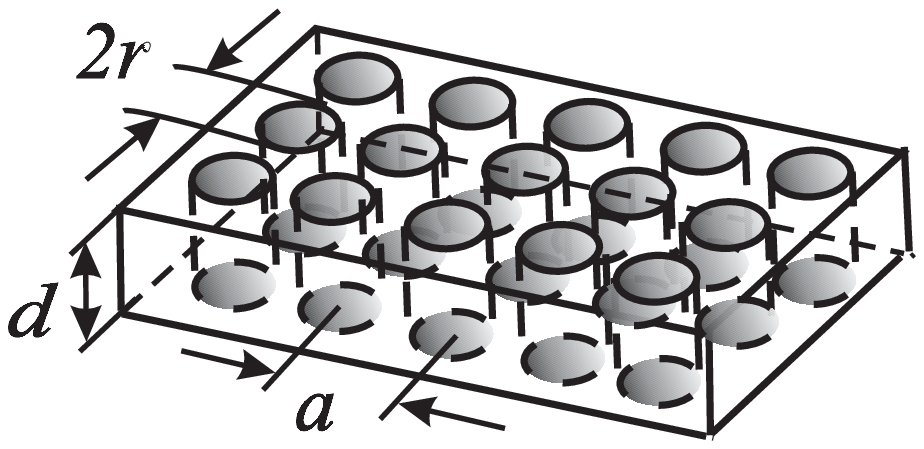,width=7.3cm,height=5.0cm}
\epsfig{figure=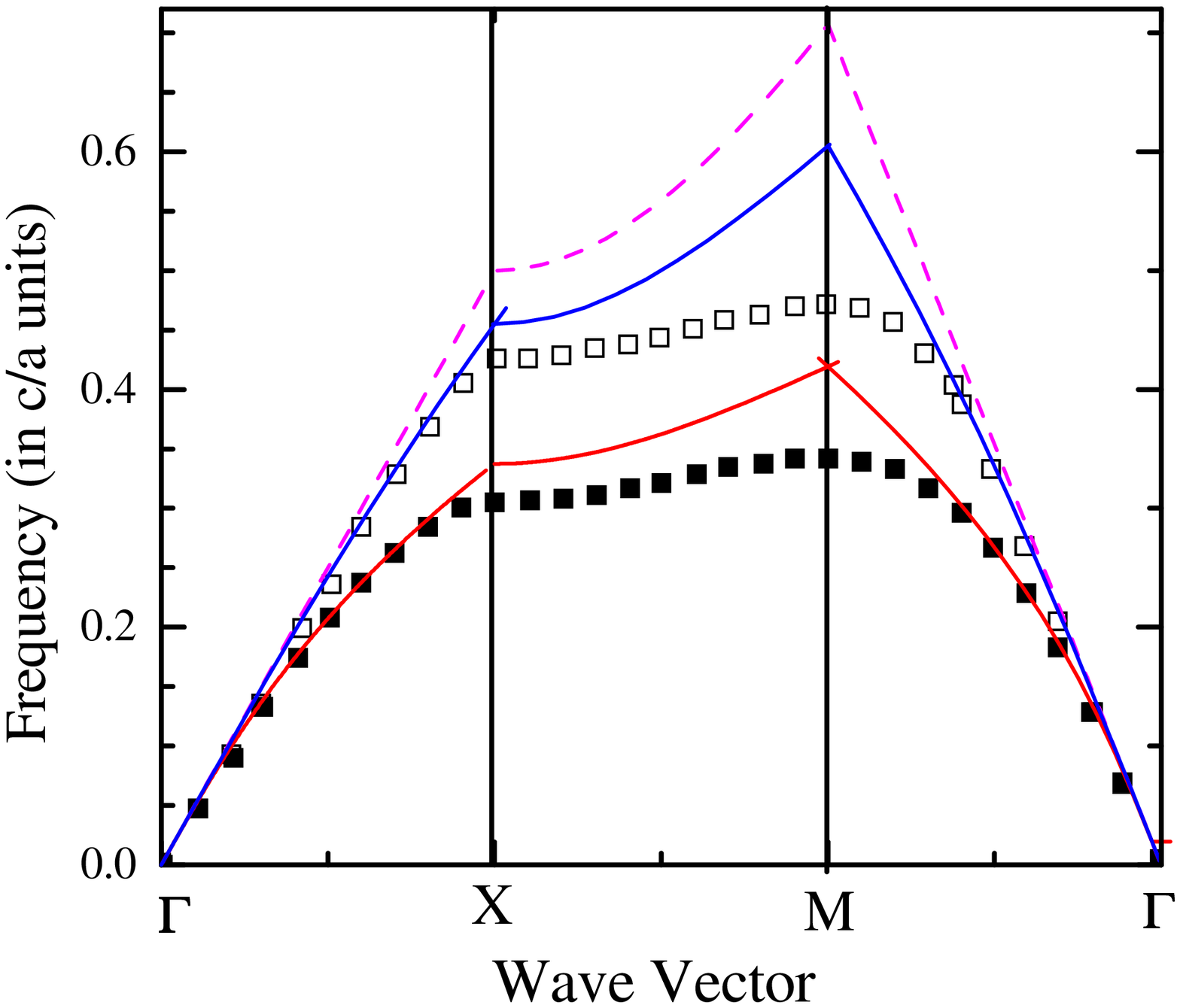,width=7.8cm,height=6.5cm} \caption[]
{Low-frequency spectra of  a photonic crystal slab composed of a
square lattice of cylindrical rods (left panel) calculated in the
``guiding power"  approximation,  Eqs. (\ref{disp-main},
\ref{kappa}). The TE mode is shown by a blue solid line, the TM mode
by red line. Results of full-scale numerical calculations
\cite{Johnson1} are shown by open (TE) and closed (TM) dots. The
dashed line indicates the light cone boundary. The filling factor is
$f=0.125$ and the cylinder's permittivity is $\epsilon_{\rm
cyl}=12$.} \label{Fig3}
\end{figure*}

The calculated spectra of TM and TE modes in the ``guiding power"
approximation are shown in Fig. 1 for a square lattice structure
with $\epsilon_{\rm cyl}=12$, $\epsilon_g=\epsilon_{b}=1$, $d/a=2$,
and $r/a=0.2$  ($n_z=0.03$). The results are compared with those of
full-scale numerical calculations \cite{Johnson1}. Our spectra shown
in Fig. 1 do not include the effect of Bragg reflection, which can
be allowed for by using a perturbative approach \cite{Sakoda}. With
Bragg reflections included \cite{SPIE}, the guiding-power spectra
approach the exact curves very closely.

Note that the guiding-power spectra are described by one parameter
$g$ for each mode. We can determine the values of these parameters
by a best fit to the exact curves. In $\pi/a$ units, the best-fit
parameters are $g_{\rm TE}=1.2$ and $g_{\rm TM}=3.2$. These values
can then be compared with those directly calculated from Eqs.
(\ref{gptm1},\ref{gpte1}), giving $g_{\rm TE, ~\rm cyl}=1.5$ and
$g_{\rm TM, ~\rm cyl}=3.2$. For TE waves, the Maxwell Garnett
approximation apparently overestimates the local field effects. If
we use the polarizability of separate cylinders, we find $g_{\rm
TE}=1.3$, which is closer to the best fit.

For the TM mode, which is of practical importance for intersubband
lasers and is our main interest in this paper, the calculated value
$g_{\rm TM, ~\rm cyl}$ provides an excellent approximation. We
regard this very good agreement as a justification for applying the
guiding power approach to calculations of the confinement factor.
Note that due to the layer anisotropy, the waveguiding for TM waves
is much stronger than that for TE waves.

Consider now the waveguiding in structures with a porous core layer.
To calculate the guiding power for TM waves in this case, we use
again Eq. (\ref{gptm0}). However, the depolarization effects are now
very different. As is well known \cite {{L&L8}}, these effects are
important when the high-permittivity component has a convex shape,
as is the case for cylinders. In contrast, for the cylindrical
holes, the effect of depolarization can be essentially neglected,
due to the concave shape of the high-permittivity component. Hence
$\epsilon_{l,~\|}$ is given by
\be \epsilon_{l,~\|}=\epsilon_{g}+(\epsilon_{b}-\epsilon_{g})f~,
\label{Eps-longp}\ee
where the fraction of high-polarizability
material in the core equals $1-f$ for a film of porosity  $f$. The
resultant TM-wave guiding power is of the form
\begin{figure*}[t] \epsfig{figure=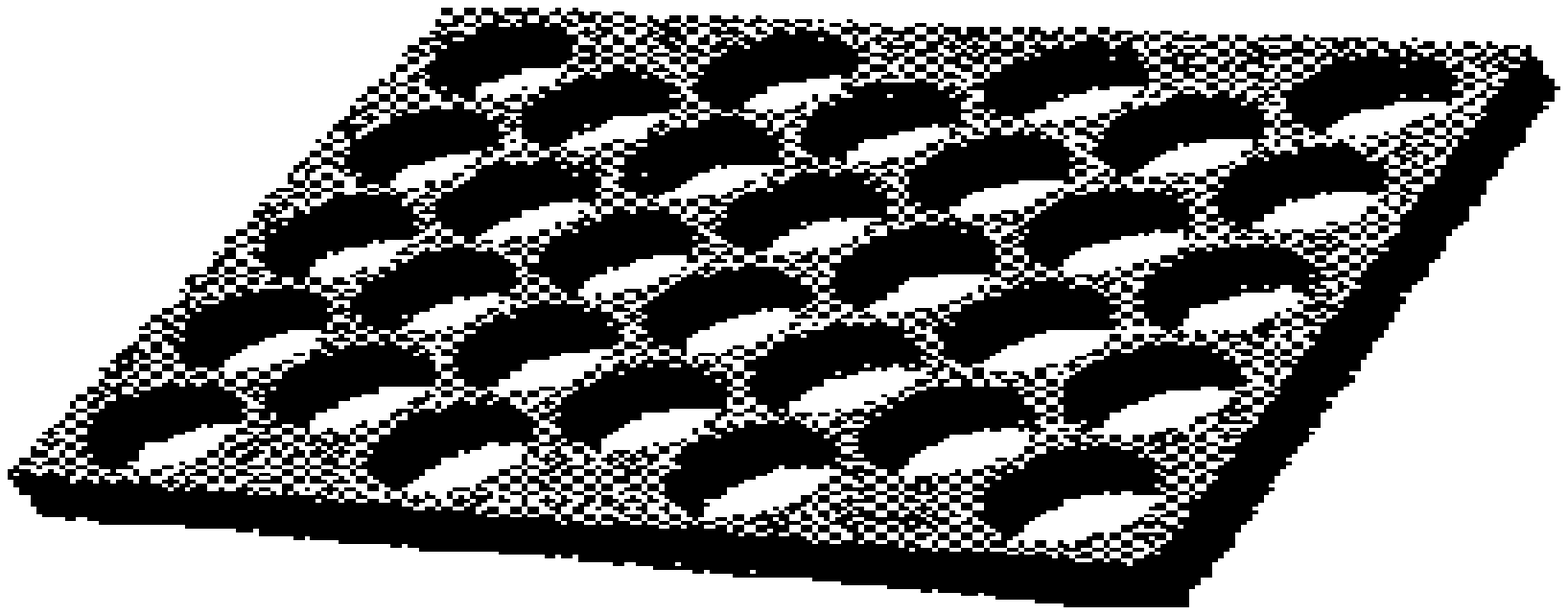,width=7.2cm,height=5.cm}
\epsfig{figure=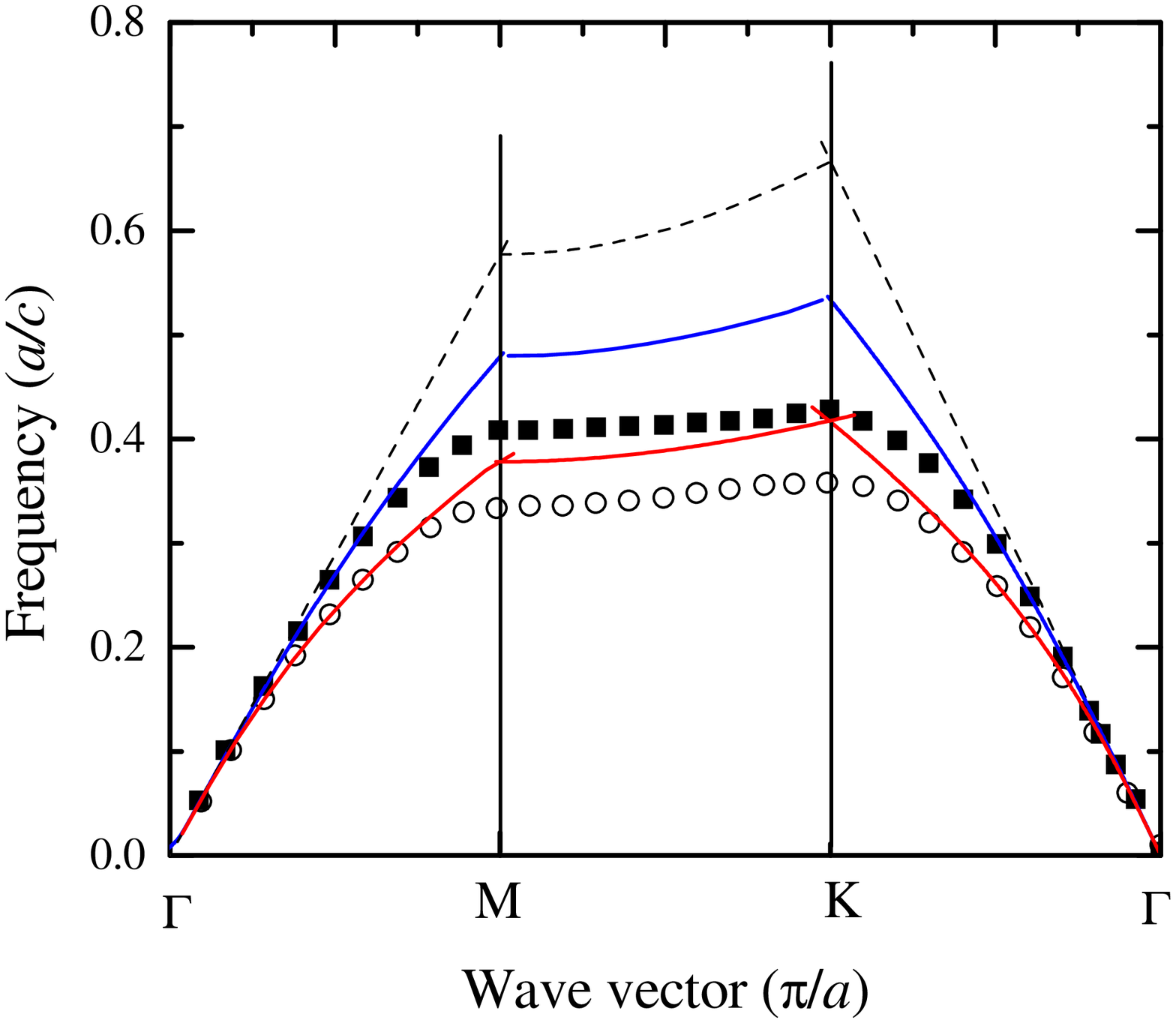,width=7.8cm,height=6.5cm} \caption[]
{Low-frequency spectra  of a photonic crystal slab formed by a
dielectric layer with a triangular lattice of air holes with lattice
constant $a$ (left panel). Layer thickness is $d=0.6a$, material
permittivity is $\epsilon_{g}=12$ and the hole radius is $r=0.45a$.
Results of the full-scale numerical calculation \cite{Johnson1} are
shown by open (for TM wave) and closed (TE) dots; red and blue lines
show the spectra of TM and TE waves in the ``guiding power"
approximation; dashed line shows the edges of the light cone.}
\label{Fig1}
\end{figure*}
\be g_{\rm TM, ~\rm por}=d\frac{(1-f)(\epsilon_{g}
-\epsilon_{b})}{\epsilon_{ b}+(1-f)(\epsilon_{g} -\epsilon_{b})}~.
\label{gptm1P}\ee
To calculate the guiding power for TE waves, we again use the
Maxwell Garnett approximation, noting, however, that it is strictly
applicable only for thick enough guiding layers, $d\gg r$ (this
restriction is, of course, compatible with $\lambda \gg d$ required
for the validity of the guiding power approximation).
For a porous layer we find
\be \epsilon_{l,~\bot}= \epsilon_g\frac{ \epsilon_b+\epsilon_g+ f
(\epsilon_b -\epsilon_{g})}{\epsilon_{b} +\epsilon_ g-f(\epsilon_b
-\epsilon_g)}~, \label{EpsBotP} \ee
whence the guiding power for TE waves is of the form
\be g_{\rm TE, ~\rm por}=d\frac{(1-f)(\epsilon_{ g}^2
-\epsilon_{b}^2)}{\epsilon_{
b}[\epsilon_{b}+\epsilon_{g}-f(\epsilon_{ b} -\epsilon_{ g})]}~.
\label{gpte1p} \ee

To check these results against those of full-scale numerical
calculations \cite{Johnson1}, we used Eqs. (\ref{disp-main},
\ref{kappa}) to calculate the TE and TM wave spectra for a
triangular lattice of pores with $d=0.6a$, and $r=0.45a$ (the
porosity $f=$0.734), see Fig. \ref{Fig1}. The values of $g$ in
$\pi/a$ units that give best fit to the spectra are $g_{\rm TE}=3.1$
and $g_{\rm TM}=1.4$. The values calculated from Eqs.
(\ref{gptm1P},\ref{gpte1p}) are $g_{\rm TE, ~\rm por}=3.4$ and
$g_{\rm TM, ~\rm por}=1.4$. As expected, the spectrum of the TM
guided wave is well described with an average dielectric constant of
the core layer over most of the Brillouin zone up to the zone
boundary, where it is strongly modified by the Bragg reflection. The
Bragg reflection can be allowed using a perturbative approach
\cite{Sakoda,SPIE}. For a high-porosity layer of fairly small
thickness, the spectrum of TE waves is also described by the guiding
power approach, but the value of $g_{\rm TE, ~\rm por}$ is
overestimated in the Maxwell Garnett approximation.

Note that both TM and TE waves remain weakly guided even in the
extreme limit of sparse structures, $a \gg d$, so long as
$\lambda \gg a$ \cite{SPIE}.

\section{Confinement factor of TM waves by a PC slab}
The modal gain in a three-layer slab waveguide with an active core
layer (ACL) of thickness $d$  can be expressed as a product of the
material gain, $G_{\rm ACL} = k_0 n''/\sqrt{\epsilon_{\rm ACL}}$ and
the dimensionless ``optical confinement factor", $\Gamma$. For
weakly guided TE waves with a smooth variation of the electric field
of the wave across the layer, the confinement factor is just
proportional to the guiding power, $\Gamma_{\rm TE}=\kappa d=
\epsilon_b k_0^2 d g_{\rm TE}/2$. For TM waves the confinement is
influenced by the weakening of the electric field in the
high-permittivity core. Generally, it can be written in the form
\cite{Visser}
\be \Gamma_{\rm TM}={{\int_{\rm ACL}E_z^2 dz} \over
\int_{-\infty}^\infty E_z^2 dz}~, \label{CF1} \ee
where the integral in the numerator is taken over the active region.
For a homogeneous core layer one can calculate the confinement
factor explicitly.  In the case of weak guiding, $\kappa d\ll 1$,
using Eq. (\ref{CF1}) and the boundary condition for the normal
component of the field at the core edge planes, we find
 \be \Gamma_{\rm TM}=
{{(k_0 d)^2}\over 2} \left( \frac {\epsilon_{b}}{ \epsilon_{g}} \right )^3 \left( 1-\frac{\epsilon_{ b}}{ \epsilon_{g}}\right) \epsilon_{g}~. \label{1}\ee
The value of $\Gamma_{\rm TM}$ is smaller than $\Gamma_{\rm TE}$ by
a factor $(\epsilon_{ b}/ \epsilon_{g})^3$. For a given core
composition (i.e. for a fixed value of $\epsilon_{ g}$) and as a
function of the cladding-layers index, $\Gamma_{\rm TM}$ has a
maximum value,
\be \Gamma_{\rm TM, ~\rm max}\approx 0.05~\epsilon_{g}(k_0 d)^2~,
\label{2}\ee
achieved when the ratio of the dielectric constants  is
$\epsilon_{b} / \epsilon_{ g}=3/4$. Note that $\Gamma_{\rm TM,~\rm
max}$ is about ten times smaller than the confinement of the TE mode
achievable in the same structure. For a fixed value of $\epsilon_{
g}$, the factor $\Gamma_{\rm TE}$ does not have a maximum, except at
$\epsilon_{b}=1$, where $\Gamma_{\rm TE}\approx 0.5~\epsilon_{\rm
g}(k_0 d)^2 \approx 10~\Gamma_{\rm TM,~\rm max}$. The difference in
confinement factors is negligible in the structures with very small
index contrast and thus very small confinement factors for both
waves \cite{ASSL}.

In the preceding section we showed that the TM waveguiding can be
enhanced and even can exceed that for the TE mode by incorporating
in the core layer cylindrical rods of high polarizability. We shall
now show that the mode confinement in the {\em active part} of the
core layer can be enhanced as well.

Consider the waveguiding in a patterned structure with initially
small difference between the dielectric constant of the active layer
(denoted by $\epsilon_g$)  and that of the cladding background,
$\epsilon_b$. Small values of $(\epsilon_g-\epsilon_b)$ are typical
for quantum cascade lasers with a multilayer active region
\cite{Capasso}. To enhance the waveguiding, we incorporate in the
structure a set of cylindrical rods of radius $r$, lattice constant
$a$, and dielectric constant $\epsilon_{\rm cyl}$. (Although we
speak of the ``lattice constant", the periodicity of rods is of no
importance here and the result can be expressed in terms of their
fill factor $f$.)
\begin{figure}
\epsfig{figure=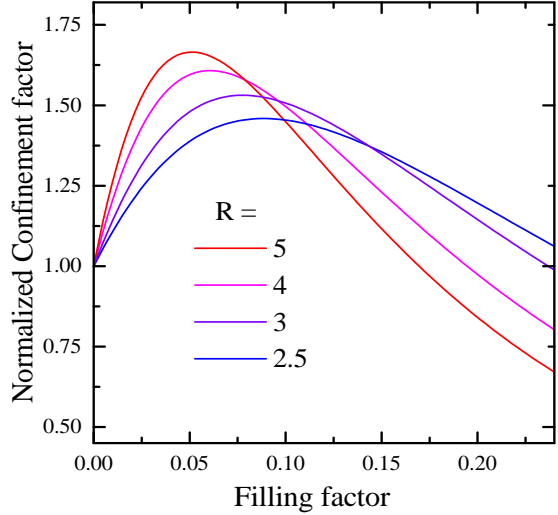,width=7.4cm,height=6.9cm} \caption[]
{Normalized confinement factor for a PC slab with a square lattice
of high-polarizability cylinders embedded in the core layer plotted
as a function of $f$ for several values of the ratio
$R=\epsilon_{\rm cyl}/\epsilon_g$, for $\epsilon_g=12$ and
$\epsilon_b=11$.} \label{Fig4}
\end{figure}
For a waveguide with a patterned core layer, we first use the
average dielectric constant to estimate the average field (which is
somewhat different from the local field), and then use Eq.
(\ref{CF1}) to calculate the confinement factor of guided TM waves
\be \Gamma_{\rm TM} ={{(k_0d)^2}\over
2}\left({\epsilon_b\over\epsilon_{l,~\|}}\right)^2\left(
1-{\epsilon_b\over\epsilon_{l,~\|}}\right) (1-f)\epsilon_g~.
\label{ConfineMG} \ee
It can be seen from Eq. (\ref{ConfineMG}) that in structures with a
high ratio $R=\epsilon_{\rm cyl}/\epsilon_b$, the enhancement of
waveguiding with $f$ due to the increasing factor
$(1-{\epsilon_b/\epsilon_{l,~\|}})$ can overwhelm at small $f$ both
the decrease in $(1-f)$ and the decreasing factor
$(\epsilon_b/\epsilon_{l,~\|})^2$ that describes reduction of the
electric field in the active layer.

The  variation of the confinement factor, calculated with Eq.
(\ref{EpsAv}) [modified to include a term proportional to
($\epsilon_g-\epsilon_b$)] and Eq. (\ref{ConfineMG}), is plotted in
Fig. \ref{Fig4}. The increase of confinement with $f$ is mainly due
to the guiding power enhanced by the better polarizability of
cylinders, hence it is most effective at small $f$. It can be seen
from Eqs. (\ref{ConfineMG}) and (\ref{EpsAv}) that the increase of
confinement with $f$ takes place only provided
$\epsilon_g>3(\epsilon_g-\epsilon_b)+1$ and $\epsilon_{\rm cyl } > 2
\epsilon_{g}$. Note, that the increase of $\epsilon_{l,~\|}$ with
$f$ towards the optimum value $\epsilon_{l,~\|,~\rm opt}$ should be
evaluated at fixed $\epsilon_b$ and $\epsilon_g$  (we have used
$\epsilon_b=11$ and $\epsilon_g=12$). This gives
$\epsilon_{l,~\|,~\rm opt}=\sqrt(3) \epsilon_{b}$. For
$\epsilon_{\rm cyl} \gg \epsilon_g$ the optimum filling factor is
small, $f \ll 1$, and the factor $1-f$ is close to unity. In this
most favorable case, the resulting confinement factor is larger than
that for the initial homogeneous structure and approaches the
optimal value given by Eq. (\ref{2}).

\begin{figure}
\epsfig{figure=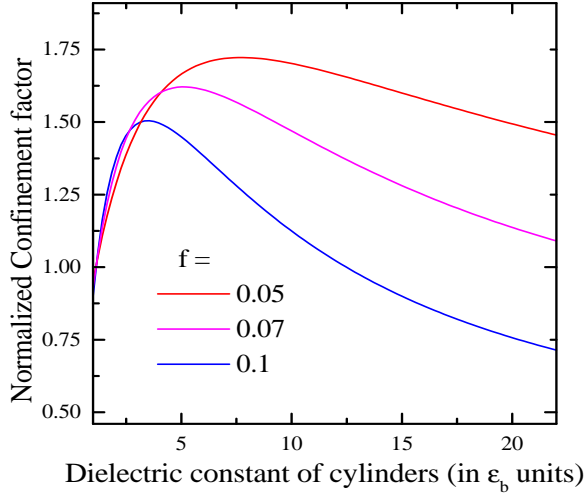,width=7.4cm,height=6.6cm} \caption[]
{Normalized confinement factor as a function of $\epsilon_{\rm cyl}$
for a PC slab with a square lattice of high-polarizability cylinders
embedded in a core layer ($\epsilon_g=12$) for several values of the
filling factor. The background permittivity outside the core layer
is $\epsilon_b=11$.} \label{Fig5}
\end{figure}

The dependence of the confinement factor on the cylinder dielectric
constant for $f=0.05$ is shown in Fig. \ref{Fig5}. In terms of the
cylinder polarizability $\alpha_z$ (Eq. \ref{PolCyl}), the
enhancement of confinement is well pronounced for $f \alpha_z> 20$,
which can be achieved in the low-wavelength limit by inclusions of
extremely highly-polarizable materials.

Next we consider the waveguiding in structures with a porous core
layer. As seen from Eqs. (\ref{gptm1P}), the guiding power of a
porous active layer is smaller than that of a homogeneous layer for
any ratio $R=\epsilon_g/\epsilon_b$ and is decreasing with $f$.
However, for large $R$ the decrease is small, while the rise of the
electric field with the layer porosity provides an enhancement of
the TM wave confinement.

\begin{figure*}[t]
\epsfig{figure=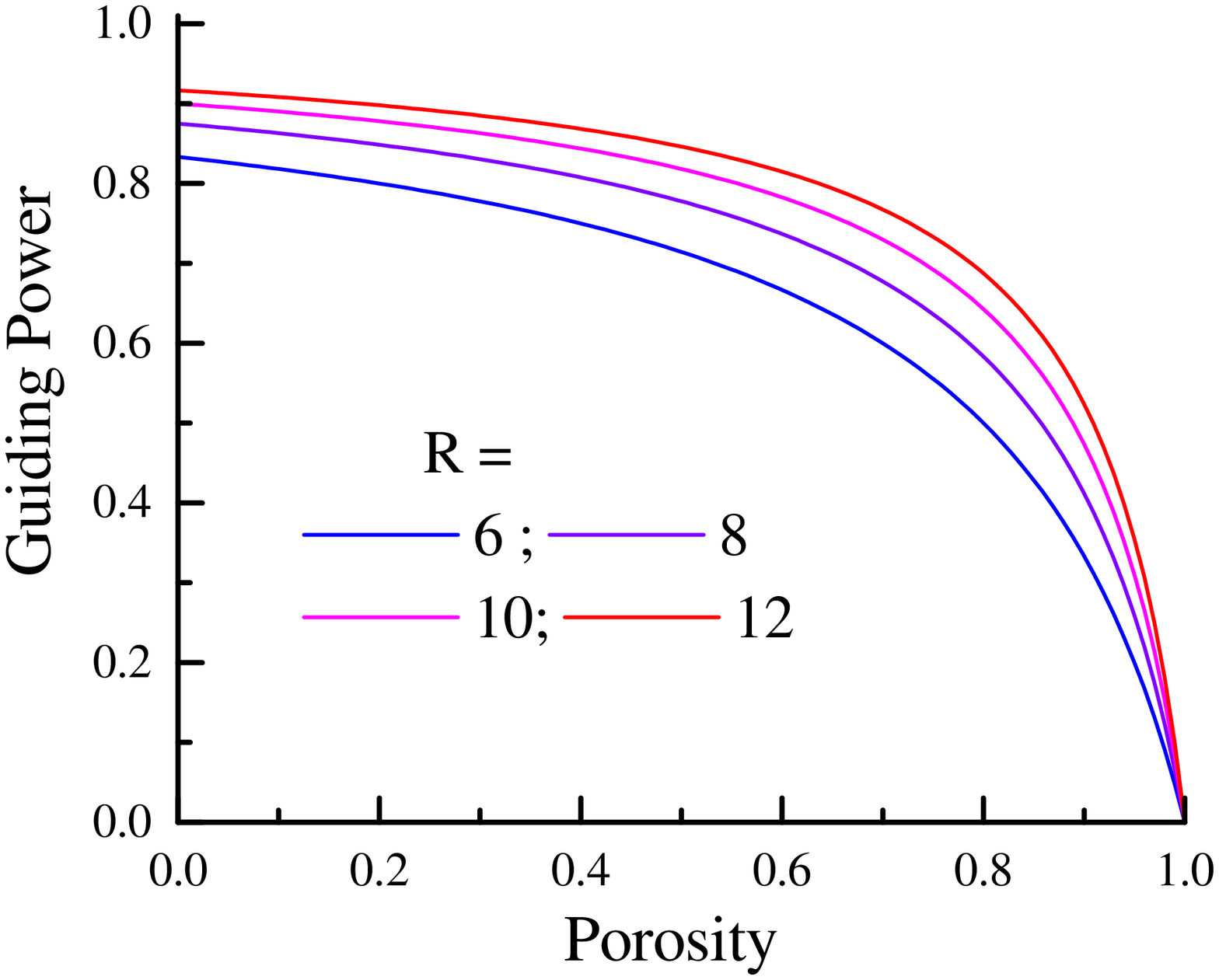,width=7.5cm,height=6.1cm}
\epsfig{figure=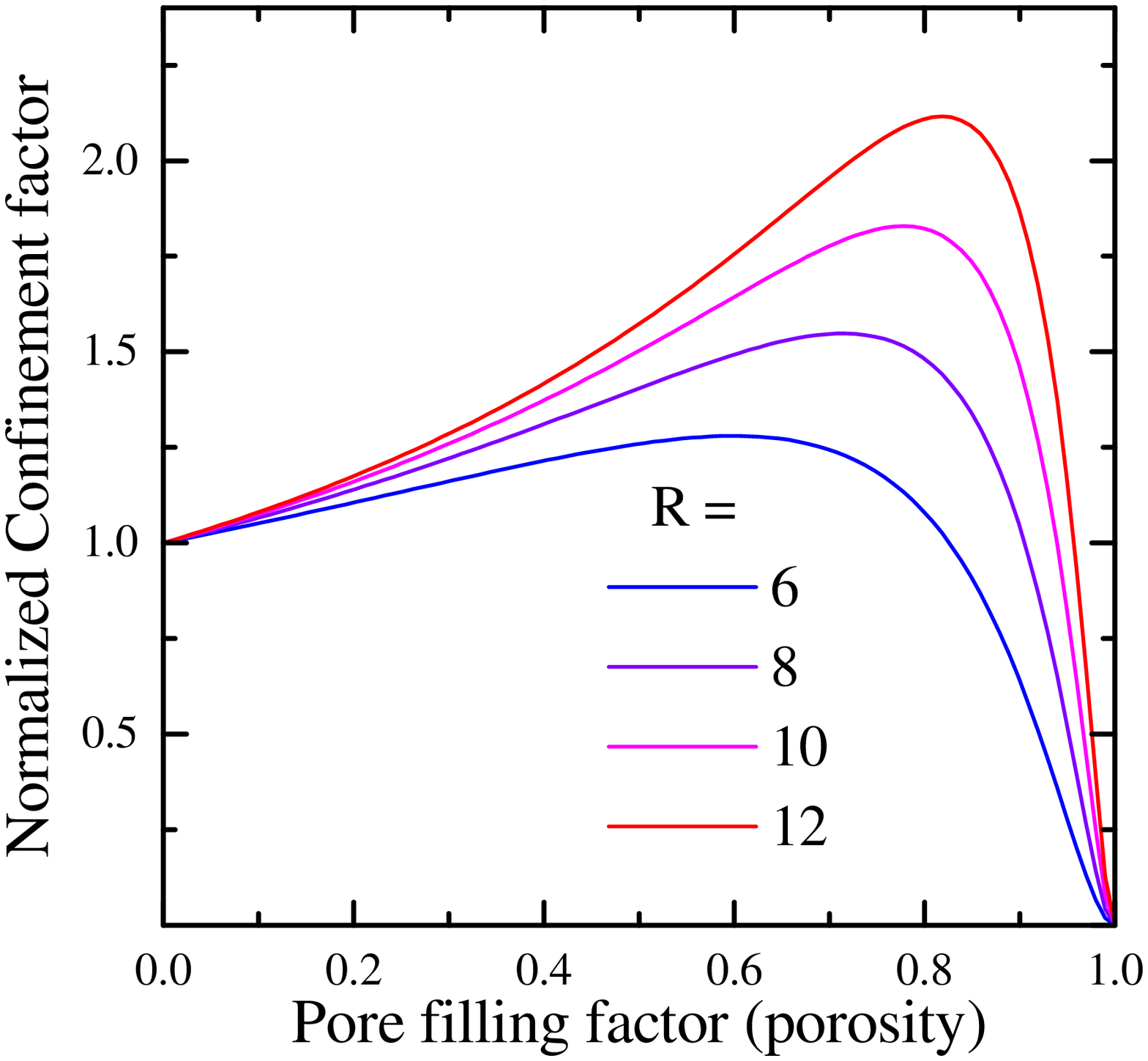,width=7.1cm,height=6.6cm} \caption[]
{Porosity dependence of the guiding power $g_{\rm TM}$ (in units of
$d$, left panel) and the normalized confinement factor $\Gamma_{\rm
TM}$ (right panel) for a PC slab with a porous core layer. Both
parameters are plotted as functions of the porosity $f$ for several
values of $R=\epsilon_g/\epsilon_b$.} \label{Fig2a}
\end{figure*}

Variations of the guiding power and the confinement factor with the
porosity $f$ for several values of the ratio $R$ are shown in Fig.
\ref{Fig2a}. For a given $R$, the maximum value of $\Gamma_{\rm TM}$
is achieved at
\be f_{\rm max}= {{R-3}\over{R-1}}~.\label{MaxFilF} \ee 
Hence, replacing the homogeneous layer by a porous layer can
increase the modal gain only using materials with $R>$ 3.
The maximum value of $\Gamma_{\rm TM}$ is
\be \Gamma_{\rm TM, ~\rm max}={{4R^3}\over{27(R-1)^2}}\Gamma_{\rm
hom}~,\label{MaxPor} \ee
where $\Gamma_{\rm hom}$ is the confinement factor for a homogeneous
core layer of the same $R$. As seen from Eq. (\ref{MaxPor}), the
increase of confinement by porosity can be tangible only for
structures with large $R$. For Si/SiO$_2$ structures with $R=6$ the
maximal confinement enhancement is achieved at $f=0.6$, but it is
not very large (1.28); for $R=12$ the enhancement is 2.1.

We remark that although the numerical values above can be determined
more accurately with more elaborate numerical calculations, the
increase of confinement by patterning initially homogeneous layers
is an exact result for waveguide structures with a high enough index
contrast between the core and the claddings. The result is based
solely on the power-law $f$ dependencies of the guiding parameter
$g_{\rm TM}$ and the field ratio inside the active layer.

\section{conclusions}

In conclusion, we considered the polarization-dependent waveguiding
of light by thin highly inhomogeneous slabs embedded in a uniform
medium. We examined exemplary slab structures comprising a monolayer
of patterned  cylindrical pores etched in a active core layer or a
pattern of high-index dielectric rods (cylinders), embedded in the
core layer. We demonstrated that for an optimal choice of the
patterned layer structure an increase of optical confinement of TM
wave is possible compared to a homogeneous layer. This increase can
be achieved both by incorporating sparsely separated high-index rods
which promote electric field penetration in the patterned structure
and by using high-porosity core layers. Our results can be useful in
the design of quantum cascade lasers.


\end{document}